\documentclass[aps,pre,12pt,reprint]{revtex4-2}
\usepackage{graphicx} 
\usepackage{amsmath}
\usepackage{hyperref}

\begin{document}
	
	\title{Spontaneous Mutations from Terahertz Proton Tunneling
	}
	\author{Noah Bray-Ali}
	\email{Email address: nbrayali@gmail.com}
	\date{November 14, 2026}
	\affiliation{Science Synergy\\ Los Angeles, CA 90045}
	\begin{abstract}
		\noindent
		Protons in the bonds between base pairs of the double helix store the code of life by breaking the dyad symmetry that swaps the sense strand with its complementary partner. When these bonds break during replication and transcription, pairs of protons switch sides, restoring dyad symmetry and destroying genetic information. The observed rate of such spontaneous mutations is obtained in the sudden approximation for bond breaking provided protons in bonds between bases tunnel across the gap with terahertz frequency.
	\end{abstract}
	
	\maketitle
	
	\section{Introduction}
	Looking down the helical axis of deoxyribonucleic acid (DNA), we find a winding stack of pairs of nucleotide bases linking the two strands of its double helix structure \cite{watson_crick1953a}. Remarkably, the molecule has a two-fold rotational symmetry axis passing through the gap between base pairs and running roughly from the minor groove of the double helix to its major groove: Half a turn about this ``dyad'' axis swaps the two helices \cite{monoclinic}. Nevertheless, the code of life breaks dyad symmetry by storing genes in the sequence of bases of the ``sense'' strand.
	
	\begin{figure}
		\begin{center}
			\includegraphics[width=3.375in,height=5.00in]{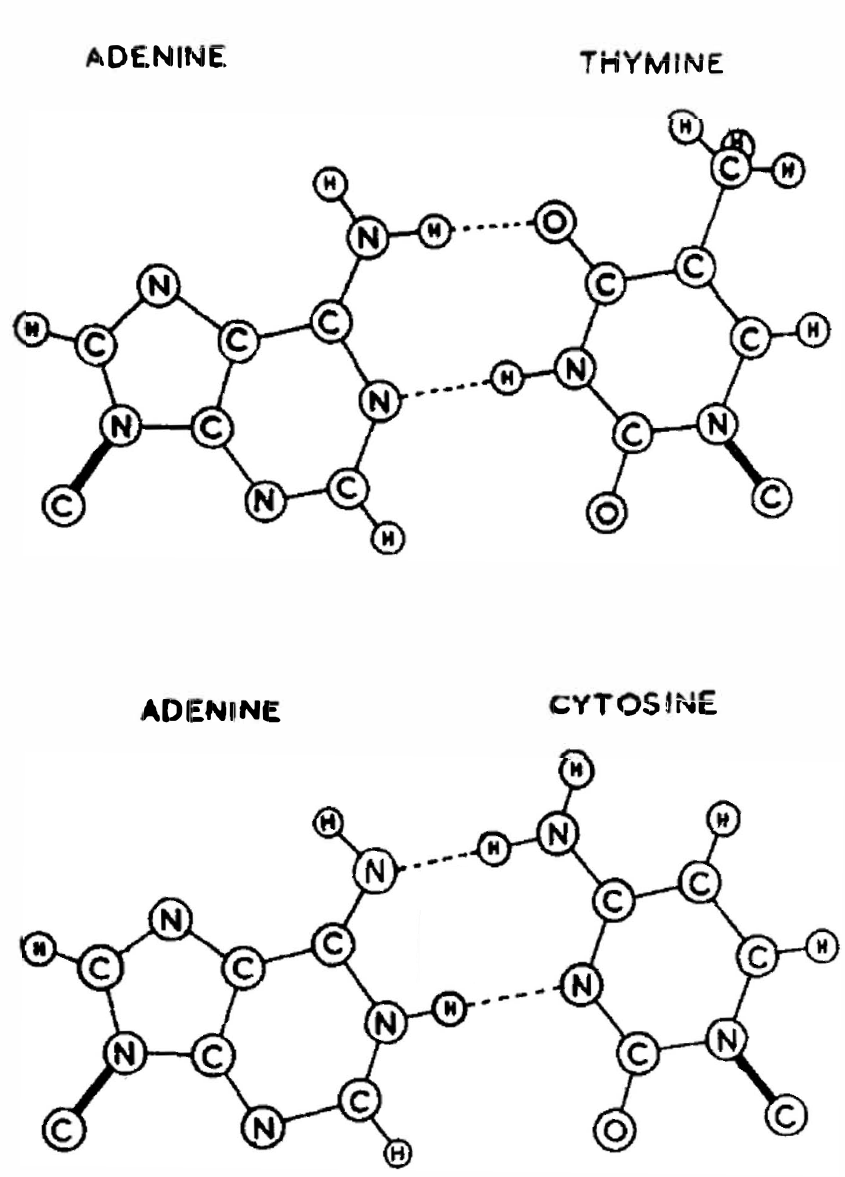}
		\end{center}
		\caption{Pairing arrangements of adenine before (above) and after (below) it has undergone tautomeric shift. The major groove of the double helix lies above the base pairs while the minor groove lies below them. We are looking down the axis of the helix. The dyad symmetry axis runs from the minor to the major groove along a line roughly half way between the bases in each pair. After the shift, the bases pair using the same atom (N for nitrogen) on either side of the bond closest to the major groove, restoring the dyad symmetry which interchanges the sense strand and its complement. (Adapted with permission from Ref. \cite{cold_spring1953}).}
		\label{shift}
	\end{figure}
	
	Spontaneous mutations restore the dyad symmetry of the double helix and let genes change by shifting the tautomeric form of bases (Fig.~\ref{shift}) \cite{cold_spring1953, lowdin,brovarets2013,slocombe2021,kim2021,slocombe2022,pushkaran2024}. Following such a shift, hydrogen bonds form between identical atoms. In wild-type bonds, by contrast, oxygen provides the lone pair of electrons and nitrogen accepts the proton in the hydrogen bond nearest the major groove. 
	
	In a mutant bond pair, a proton pair swaps sides like the electron pair in the covalent chemical bond \cite{pauling}. And just as the covalent bond is spin symmetric, so also the mutant hydrogen bond pair is dyad symmetric. When we split the mutant pair, there is an equal chance of finding either tautomeric form on a given base provided only that the other base has the complementary form. 

	The chance of spontaneous mutation, $P=2.0\times10^{-10}$, gives roughly one tautomeric shift for every few billion base pairs that are replicated or transcribed \cite{mute_rate,drake1991,wu2017}. In the sudden approximation, the wild-type proton pair wavefunction for a base pair between guanine $G$ and cytosine $C$ then must take the following form \cite{dirac,born},
	\begin{equation}
		\left|W\right\rangle=\left|\overline{GC}\right\rangle+\epsilon\left|\overline{G^*C^*}\right\rangle.
		\label{proton_pair}
	\end{equation}  
	Here $\left|\overline{GC}\right\rangle$ is the effective dominant tautomeric form (See Eq.~\ref{eq:overline-AT}), $\left|\overline{G^*C^*}\right\rangle$ has tautomeric shifts in both bases (See Eq.~\ref{eq:overline-A*T*}), and $\epsilon=\sqrt{2P}=2.0\times10^{-5}$ is the probability amplitude for these shifts to take place. 
	\begin{table}
		\begin{tabular}{|c|c|c|c|}
			\hline
			Parameter&Frequency (THz) & Energy (meV)& Period (fs)\\
			\hline
			$t$&0.48&2.0&2100\\
			\hline
			$\Delta$&110&470&8.8\\
			\hline
			$U$&$210$&$870$&$4.8$\\
			\hline
			$J$&$2.3\times10^{-3}$&$9.4\times10^{-3}$&$0.44\times10^6$\\
			\hline
		\end{tabular}
		\caption{Proton parameters governing spontaneous mutation. Proton tunneling within hydrogen bonds between nucleotide bases in DNA occurs at the terahertz frequency, $t/h=0.48~{\rm THz}$ \cite{planck}, set by the proton tunneling energy-scale, $t$ \cite{lowdin,slocombe2021,kim2021}. The energy splitting, $\Delta$, separates the dominant and rare tautomeric forms of nucleotide bases \cite{delta}, and gives a rough sense of the femtosecond time-scale, $h/\Delta=8.8~{\rm fs}$ \cite{planck}, for hydrogen bond-breaking between bases during DNA replication and transcription \cite{kornberg1994}. Charge-transfer energy, $U$, provides the high-temperature scale, $U/k=10000~{\rm K}$ \cite{delta_u}, protecting the coherence of the ground-state from thermal fluctuations within the living cell \cite{what-is-life}. The emergent microwave-frequency energy-scale, $J=2t^2/U=h\times2.3~{\rm GHz}$, gives the nano-second time-scale, $h/J=0.44~{\rm ns}$, for the correlated proton pair swaps that create spontaneous mutations.} 
		\label{param_table}
	\end{table}

	During replication and transcription, the base pair splits on a femtosecond time-scale set, according to the uncertainty principle \cite[pgs.~150--153]{landau}, by the energy needed to break the hydrogen bonds, roughly $0.75~{\rm eV}$ for $GC$ \cite{kornberg1994}, which is quick compared to the picosecond-long time-scale for tautomeric shifts due to terahertz-frequency proton tunneling between bases \cite{brovarets2013,slocombe2021,slocombe2022}. From the wild-type wavefunction in Eq.~(\ref{proton_pair}), the chance of finding the shifted form is then simply, $\epsilon^2=2P$, the square of the probability amplitude for the shift \cite{born}. This is twice the chance of spontaneous mutation, $P$, since the shifted guanine, $G^*$, has, by dyad symmetry, an equal chance of being found in either tautomeric form once it binds to thymine, $T$, yielding the mutant wavefunction, 
	\begin{equation}
		\left |M \right\rangle=\frac{1}{\sqrt{2}}\left| \overline{G^*T}\right\rangle+\frac{1}{\sqrt{2}}\left| \overline{GT^*}\right\rangle.
		\label{mutant_wave}
	\end{equation}
	
	In this Article, we show that terahertz-frequency proton tunneling between base pairs creates the wild-type and mutant proton pair wavefunctions responsible for spontaneous mutation, as envisaged by L\"{o}wdin \cite{lowdin}. The Methods section shows how to express the wild-type tautomeric shift probability amplitude, $\epsilon=2t^2/(U\Delta)$, in terms of the energy-scale, $t=h\times0.48~{\rm THz}=2.0~{\rm meV}$ \cite{planck,slocombe2021,kim2021}, for this proton tunneling, the energy splitting between tautomers, $\Delta=h\times110~{\rm THz}=470~{\rm meV}$ \cite{delta}, and the charge-transfer energy, $U=h\times210~{\rm THz}=870~{\rm meV}$ \cite{delta_u}, for putting both protons on the same base. The Results section justifies this straightforward ``closed'' quantum mechanical analysis in terms of the high temperature scale, $U/k=10000~{\rm K}$, which protects the coherence and entanglement of both the wild-type and mutant proton-pair wavefunctions against thermal fluctuations at room temperature, $300~{\rm K}$, following an insight articulated by Schr\"{o}dinger \cite{what-is-life,zwitterion}.
	
	Table \ref{param_table} summarizes the proton parameters governing spontaneous mutation in the living cell. The rate of genetic change from spontaneous mutation follows from the emergence of the radio-frequency scale, $J=2 t^2/U=h\times2.3~{\rm GHz}$, for the correlated proton pair swaps created by terahertz frequency proton tunneling. We end with Discussion of the meaning of these results for evolution and for aging \cite{lowdin}.
	
	
	
	\section{Methods}
	The wild-type proton pair wavefunction, $\left|W\right\rangle$, is the lowest energy state of the effective Hamiltonian \cite{dirac,schrodinger},
	\begin{eqnarray}
		H_{\rm eff}=&-&\frac{\Delta}{2}\left(\left|\overline{GC}\right\rangle\left\langle \overline{GC}\right|-\left|\overline{G^*C^*}\right\rangle\left\langle \overline{G^*C^*}\right|\right)\nonumber\\
		&-&J\left(\left|\overline{GC}\right\rangle\left\langle \overline{G^*C^*}\right|+\left|\overline{G^*C^*}\right\rangle\left\langle \overline{GC}\right|\right),
		\label{effective}
	\end{eqnarray}
	where the first term splits the tautomers and the second term swaps the protons \cite{lowdin}. The swap energy, $J=2t^2/U=9.4\times10^{-3}~{\rm meV}$, is small compared to the tautomeric splitting, $\Delta=470~{\rm meV}$ \cite{delta}. This means we can find the wild-type wavefunction, $\left|W\right\rangle=\left|\overline{GC}\right\rangle+\epsilon\left|\overline{G^*C^*}\right\rangle$, by time-independent first-order perturbation theory \cite[pg.~133--139]{landau}.
	
	We simply act on the proton pair wavefunction, $\left|W\right\rangle$, in Eq.~(\ref{proton_pair}), with the effective Hamiltonian, in Eq.~(\ref{effective}), and keep only the terms that are first-order in the small quantities, $J/\Delta=2t^2/(U\Delta)=2\times10^{-5}$, and, $\epsilon$ \cite{dirac},
	\begin{equation}
		H_{\rm eff}\left|W\right\rangle=-(\Delta/2)\left|\overline{GC}\right\rangle+(\epsilon\Delta/2-J)\left|\overline{G^*C^*}\right\rangle.
		\label{energy_one}
	\end{equation}
	Then, we use the fact that the energy of the proton pairs in the ground-state wavefunction does not change to first-order in the small quantity, $J/\Delta$, from the value, $-\Delta/2$, that the pair would have without proton swaps \cite{schrodinger},
	\begin{equation}
		-(\Delta/2)\left|W\right\rangle=-(\Delta/2)\left|\overline{GC}\right\rangle-(\epsilon\Delta/2)\left|\overline{G^*C^*}\right\rangle.
		\label{energy_two}
	\end{equation}
	Matching the right-hand sides of Eq.~(\ref{energy_one}) and Eq.~(\ref{energy_two}), we find, at last, the result for the probability amplitude for proton swap, $\epsilon=J/\Delta=2\times10^{-5},$ which is small enough that we can simply stop at first-order in perturbation theory for the wild-type wavefunction.
	
	Terahertz proton tunneling generates the effective Hamiltonian, $H_{\rm eff}$, at second-order in time-independent perturbation theory \cite[pgs.~133--139]{landau} from the full Hamiltonian matrix \cite{matrix}, 
	\begin{equation}
		H=\begin{pmatrix}
			-\Delta/2 & 0 & -t & -t\\
			0 & \Delta/2 & -t & -t \\
			-t & -t & U & 0 \\
			-t & -t & 0 & U 
		\end{pmatrix}
		\label{bare}
	\end{equation}
	where the matrix acts on the column vector, $(a,b,c,d)^T$, for the proton pair wave function, $a\left|GC\right\rangle+b\left|G^*C^*\right\rangle+c\left|G^+C^-\right\rangle+d\left|G^-C^+\right\rangle$.  The zwitterion states, $\left|G^+C^-\right\rangle$, and, $\left|G^-C^+\right\rangle$, have charge-transfer energy, $U=870~{\rm meV}$ \cite{delta_u}, compared to the tautomers, $\left|GC\right\rangle$, and, $\left|G^*C^*\right\rangle$. The zwitterion states arise from the tautomers when a single proton hops within a hydrogen bond between nucleotide bases \cite{lowdin,pushkaran2024}.
	
	The base on which the proton lands has positive charge from the extra proton. At the same time, the base from which the proton came is left with negative charge from the extra lone pair of electrons. The small size of the tunneling energy, $t=U/440$, compared to the charge-transfer energy, $U$, means we can effectively eliminate the high-energy zwitterion states from the low-energy tautomeric physics through a simple change of basis.
	
	In the language of quantum optics, the zwitterions have a ``bright'' state, $\left|+\right\rangle$, that mixes with the tautomers and a ``dark'' state, $\left|-\right\rangle$, that does not mix with them \cite{bright_dark}. The zwitterion dark state has equal probability to land in either zwitterion but opposite probability amplitude:
	\begin{equation}
		\left|-\right\rangle=\frac{1}{\sqrt{2}}\left(\left|G^+C^-\right\rangle-\left|G^-C^+\right\rangle\right)
	\end{equation}
	It retains the energy, $U$, that it would have without proton tunneling \cite{dark_state}. Destructive interference from the two zwitterions decouples the dark state from the tautomers. 
	
	Meanwhile, the zwitterion bright state mixes with the tautomers and has equal probability amplitude for both zwitterions.
	\begin{eqnarray}
		\left|+\right\rangle&=&-\frac{\sqrt{2}t}{U}\left(\left|GC\right\rangle+\left|G^*C^*\right\rangle\right)\nonumber\\
		&+&\frac{1}{\sqrt{2}}\left(\left|G^+C^-\right\rangle+\left|G^-C^+\right\rangle\right)
	\end{eqnarray}
	The zwitterion bright state, $\left|+\right\rangle$, lowers its energy to, $U-2J$, through constructive interference, where, $J=2t^2/U=h\times2.3~{\rm GHz}$, is the microwave energy-scale for proton pair swaps \cite{bright_state}. This new scale emerges from the terahertz-frequency $t=h\times0.48~{\rm THz}$ for proton tunneling when we change basis to the zwitterion bright and dark states.
	
	Coming back to the tautomers, the dominant and rare forms mix with the zwitterion states to make a new pair of tautomers whose dynamics are governed by the effective Hamiltonian in Eq. (\ref{effective}). The role of the dominant tautomer in the effective Hamiltonian is played by the state, 
	\begin{equation}
		\label{eq:overline-AT}
		\left|\overline{GC}\right\rangle=\left|GC\right\rangle+(t/U)\left(\left|G^+C^-\right\rangle+\left|G^-C^+\right\rangle\right).
	\end{equation}
	It is easy to check that this state has no overlap with the bright and dark zwitterion states \cite{overlap}. Similarly, for the rare tautomer, we must take the following state,
	\begin{equation}
		\label{eq:overline-A*T*}
		\left|\overline{G^*C^*}\right\rangle=\left|G^*C^*\right\rangle+(t/U)\left(\left|G^+C^-\right\rangle+\left|G^-C^+\right\rangle\right), 
	\end{equation}
	to ensure it too has no overlap with the states of the new zwitterion basis, $\left|\pm\right\rangle$. 
	
	With this simple change of basis, the zwitterion bright and dark states split off from the radio-frequency dynamics of the tautomers relevant for spontaneous mutation. The full Hamiltonian simplifies to the new matrix \cite{matmult},
	\begin{equation}
		V^THV=\begin{pmatrix}
			-\Delta/2 & J & 0 & 0\\
			J & \Delta/2 & 0 & 0\\
			0 & 0 & U-2J & 0 \\
			0 & 0 & 0 & U 
		\end{pmatrix}
		\label{clothed}
	\end{equation}
	where $V$ (and  $V^T)$ are the matrices whose columns (rows) are the vectors that give the wavefunctions for the new basis states in terms of the original basis. The matrix now acts on column vectors that give the proton pair wavefunction in the new basis of states. The new basis is simply the new dominant tautomer, $\left|\overline{GC}\right\rangle$, the new rare tautomer, $\left|\overline{G^*C^*}\right\rangle$, the bright zwitterion, $\left|+\right\rangle$, and the dark zwitterion, $\left|-\right\rangle$. 
	
	The effective Hamiltonian from Eq.~(\ref{effective}) appears in the upper left corner and creates proton swaps that exchange the new tautomer basis states, as envisaged by L\"{o}wdin \cite{lowdin}. Meanwhile, the zwitterion bright and dark states are stationary states of the full Hamiltonian with energies, $U-2J$, and, $U$, respectively. The change of basis and the new form for the full Hamiltonian in that basis are exact up to and including terms of order, $(t/U)^2=5.3\times10^{-6}$. 
	
	The energy, $t=U/440$, gained by terahertz proton tunneling between base pairs is small compared to the near-infrared-scale energy, $U=870~{\rm meV}$ \cite{delta_u}, needed to accommodate the charge transfer. This means we can simply stop at second-order in the time-independent perturbation theory for the effective Hamiltonian of the tautomeric shifts that generate spontaneous mutation \cite{zwitterion}. The resulting effective Hamiltonian in Eq.~(\ref{effective}) generates proton pair swaps with the emergent microwave frequency, $J=2t^2/U=h\times2.3~{\rm GHz}$.  
	
	\section{Results}
	The chance of spontaneous mutation, $P=(J/\Delta)^2/2=2.0\times10^{-10}$, emerges from the wild-type proton pair wavefunction, $\left|W\right\rangle$, of nucleotide base pairs in DNA given in Eq. (1). The charge-transfer energy, $U=k\times 10000~{\rm K}$ \cite{delta_u}, protects the coherence and entanglement of both the wild-type and mutant proton-pair wavefunctions against thermal fluctuations at room temperature $300~{\rm K}$ \cite{what-is-life, zwitterion}. Meanwhile, the wavefunction has amplitude, $J/\Delta$, to swap protons between the bases, where, $J=2t^2/U=h\times2.3~{\rm GHz},$ is the microwave-frequency energy-scale that emerges from the terahertz-frequency proton tunneling amplitude, $t=h\times0.48~{\rm THz}$ \cite{planck}, and the charge-transfer energy, $U=870~{\rm meV}$ \cite{delta_u}. 
	
	When bonds between bases break during replication and transcription, the femtosecond time-scale for bond-breaking, on the order of the time-scale, $h/\Delta=8.8~{\rm fs}$ \cite{kornberg1994}, set by the tautomer energy-splitting \cite{delta}, is fast compared to the picosecond time-scale needed for protons to tunnel between bases at the terahertz proton tunneling frequency, $t=h\times0.48~{\rm THz}=h/(2.1~{\rm ps})$ \cite{brovarets2013,slocombe2021,slocombe2022}. Similarly, the nearly nanosecond time-scale required for correlated pair swaps at the emergent microwave-frequency scale, $J=2t^2/U=h/(4.8~{\rm ns})$, is much slower than the time-scale for bond-breaking. In the sudden approximation to the bond-breaking process, the protons then land in the rare tautomer form with probability, $\epsilon^2=(J/\Delta)^2=4\times10^{-10}$ \cite{born}, given by the square of the probability amplitude, $\epsilon$, for the correlated proton pair swap within the wild-type proton pair wavefunction at frequency, $J/h=2.3~{\rm GHz}$, between the rare and dominant tautomers with energy splitting, $\Delta=470~{\rm meV}$ \cite{delta}. 
	
	The tautomeric shift, induced by bond-breaking, forces the new base pairs that form during replication and transcription to have the mutant wavefunction, $\left|M\right\rangle$, given in Eq. (2). The mutant wavefunction has equal amplitude for a given base to be found in either tautomeric form. This restores the dyad symmetry between the two strands of the double helix and destroys the genetic information that was stored within the old base pair.
	
	The physical picture for the proton swap proceeds in two coherent steps \cite{brovarets2013,slocombe2022}. First, the proton closer to the major groove, say, tunnels across the gap between bases. This costs charge-transfer energy but is not forbidden by any symmetry principle. Next, the other proton, located closer to the minor groove, tunnels across the gap to gain back the charge transfer energy. At the end of the coherent second-order proccess, the protons have swapped. 

	\section{Discussion}
	Terahertz proton tunneling in the hydrogen bonds between nucleotide base pairs within DNA generates radio-frequency proton pair swaps that drive spontaneous mutation, as envisaged by L\"{o}wdin \cite{lowdin}. Yet both processes are slow compared to the femtosecond-scale bond-breaking during replication and transcription. The sudden splitting makes the pair choose at random which tautomer to take: the low-energy form found in free nucleotides within the cytoplasm, or the high-energy form that generates mutations. Remarkably, the chance they land in the rare tautomer gives roughly twice the observed spontaneous mutation rate with the factor of two reflecting the restoration of dyad symmetry between the two strands of the double helix structure when the rare tautomer binds to a free nucleotide during replication.
	
	The spontaneous mutation rate acts as a molecular clock driving evolution. Each generation must replicate its genome to pass along genetic information to its descendants. However, the molecular process of replication itself generates changes in the sequence of bases within DNA. These spontaneous mutations restore the symmetry between the sense strand and its complementary partner. They destroy genetic information at random within the genome at a regular rate. The spontaneous mutation rate can be expressed in terms of the energy splitting between tautomers and the emergent energy-scale for proton pair swap between nucleotides.
	
	Spontaneous mutation also sets the life span of organisms by making cells age \cite{lowdin}. Each time the cell uses the code of life to live its life, it must break bonds between base pairs to read the base sequence on the sense strand. Like replication, the transcription process itself generates spontaneous mutations. The sudden splitting of the strands makes base pairs choose at random which tautomer to take. 
	
	During transcription, when a base pair in DNA lands in the rare form, the base on the sense strand then must form a mutant bond with the ``wrong'' base in the complementary messenger ribonucleic acid (mRNA) molecule. This mutant bond then breaks to release the single-stranded mRNA. By the dyad symmetry created within the mutant bond, though, the base on the sense strand of the DNA has an even chance of coming back as either the dominant or the rare tautomer. 
	
	To complete transcription, the sense strand base returns to bind its partner. If it returns in the dominant form, however, the bases do not bind: Both bases now have their protons close to the same groove (major or minor) of the double helix and the electrostatic repulsion between protons keeps the bases apart. To bind the strands, the transcription machinery must remove one of the bases and replace it with a free nucleotide in the dominant form. 
	
	Since the sense strand makes and breaks twice as many bonds as its partner does, the machinery prefers to use the complementary base as its template for repair. This locks in the spontaneous mutation as a mutant bond with dyad symmetry between the two strands and complete loss of genetic information. In this way, the act of using the code of life to live in fact destroys the code itself, creating an aging of the genetic code within the living cell that sets a fundamental limit on the life span of a given generation. 
	
	
	By way of conclusion, it is worth recalling the key steps. The analysis began by recognizing that protons tunnel with terahertz frequency between the nucleotides in the base pairs that bind the double helix structure of DNA and that store the code of life. The proton that hops back across the gap, however, need not be the same one that hopped first. The resulting proton swap between base pairs leaves both bases in the pair in a rare, high-energy tautomer form. The emergent radio-frequency proton pair swap energy-scale competes with the tautomer energy splitting to set the spontaneous mutation rate. During replication and transcription, the hydrogen bonds between base pairs break fast and force the base pair to choose which form to take. This random choice drives the genetic drift between generations and leads to evolution. It also leads to the aging of organisms and sets the length of each generation. 
	
	\section*{Acknowledgements}
	\noindent
	Takeshi Irie, Warren Ruder, and Alan Heins provided helpful suggestions, constructive criticism, and insightful comments. Carol Miu inspired, encouraged, and participated in the deep-dive into the history of the August 1951 discovery of the dyad symmetry of DNA by Rosalind Franklin in conversation with Dorothy Hodgkin. Discussions with James Boedicker, Sam Peng, and Ian Spielman were invaluable in shaping and sharpening the argument for the relevance to the conditions actually found in the living cell. Daria Mazura provided constant and critical support without which the work reported here would not have been possible. 



\begin{thebibliography}{99}
\bibitem{dark_state} We start by writing the zwitterion dark state as a column vector in the same basis, $(\left|GC\right\rangle,\left|G^*C^*\right\rangle,\left|G^+C^-\right\rangle,\left|G^-C^+\right\rangle)$, used to express the Hamiltonian matrix in Eq.~(\ref{bare}) \cite{matrix}, 
		\begin{eqnarray}
			\left|-\right\rangle&=&\frac{1}{\sqrt{2}}\left(\left|G^+C^-\right\rangle-\left|G^-C^+\right\rangle\right)\nonumber\\
			&=&\frac{1}{\sqrt{2}}(0,0,1,-1)^T.
		\end{eqnarray}
		Then, we simply multiply the resulting column vector by the Hamiltonian matrix in Eq.~(\ref{bare}) using the matrix multiplication rules \cite{matrix},
		\begin{eqnarray}
			H|-\rangle=U\frac{1}{\sqrt{2}}(0,0,1,-1)^T=U|-\rangle.
		\end{eqnarray}
		Finally, we interpret the result to mean that the zwitterion dark state, $|-\rangle$ has energy $U=870~{\rm meV}$ \cite{delta_u}, given by the energy cost of having two protons on one base and no protons on the other base \cite{matrix}.
\bibitem{bright_state} We start by writing the zwitterion bright state as a column vector in the same basis $(\left|GC\right\rangle,\left|G^*C^*\right\rangle,\left|G^+C^-\right\rangle,\left|G^-C^+\right\rangle)$, used to express the Hamiltonian matrix in Eq.~(\ref{bare}) \cite{matrix} 
		\begin{eqnarray}
			\left|+\right\rangle&=&-\frac{\sqrt{2}t}{U}\left(\left|GC\right\rangle+\left|G^*C^*\right\rangle\right)\nonumber\\
			&+&\frac{1}{\sqrt{2}}\left(\left|G^+C^-\right\rangle+\left|G^-C^+\right\rangle\right)
			\nonumber\\
			&=&\left(-\frac{\sqrt{2}t}{U},-\frac{\sqrt{2}t}{U},\frac{1}{\sqrt{2}},\frac{1}{\sqrt{2}}\right)^T.
		\end{eqnarray}
		Then, we simply multiply the resulting column vector by the Hamiltonian matrix in Eq.~(\ref{bare}) using the matrix multiplication rules \cite{matrix}
		\begin{eqnarray}
			H|+\rangle&=&(U-2J)\times\left(-\frac{\sqrt{2}t}{U},-\frac{\sqrt{2}t}{U},\frac{1}{\sqrt{2}},\frac{1}{\sqrt{2}}\right)^T\nonumber\\
			&=&(U-2J)\left|+\right\rangle,
		\end{eqnarray}
		where $J=2t^2/U$ is the emergent energy-scale for correlated proton pair swaps. Finally, we interpret the result to mean that the zwitterion bright state, $\left|+\right\rangle$, lowers its energy in the amount, $2J=2\times2\times(2.0~{\rm meV})^2/(870~{\rm meV})=19\times10^{-3}~{\rm meV}$, by mixing with the tautomers \cite{schrodinger,matrix}.
\bibitem{matmult} The matrix product, $MV=(Mv_1, Mv_2, \ldots, Mv_N)$, of matrix, $M$, and matrix, $V=(v_1,v_2,\ldots,v_N)$, is simply the matrix whose column vectors, $Mv_1, Mv_2,\ldots,Mv_N$, come from acting with matrix, $M$, on the column vectors, $v_1, v_2,\ldots v_N$, of matrix $V$. The change of basis matrix, $V$, in Eq.~(\ref{clothed}) takes the form \cite[pgs.~32--32]{landau},
		\begin{equation}
			V=\begin{pmatrix}
				1 &0  &  -\sqrt{2}t/U&0 \\
				0&  1&  -\sqrt{2}t/U&  0\\
				t/U& t/U & 1/\sqrt{2} &1/\sqrt{2}  \\
				t/U&  t/U& 1/\sqrt{2} & -1/\sqrt{2} 
			\end{pmatrix}
			\label{bare}
		\end{equation}    
		where the columns of $V$ are the column vectors that give the states of the new basis (dominant tautomer, rare tautomer, bright zwitterion, dark zwitterion) in terms of the old basis, ($\left|GC\right\rangle, \left|G^*C^*\right\rangle, \left|G^+C^-\right\rangle, \left|G^-C^+\right\rangle)$.
\end{thebibliography}
\end{document}